\documentclass[twocolumn,showpacs,superscriptaddress,prb]{revtex4}

\usepackage{graphicx}% Include figure files
\usepackage{dcolumn}% Align table columns on decimal point
\usepackage{bm}% bold math

\begin{document}
\preprint{APS/123-QED}

\title{Suppression of ferromagnetism in the LaV$_x$Cr$_{1-x}$Ge$_3$ system}

\author{Xiao Lin}
 \author{Valentin Taufour}

\affiliation{Department of Physics and Astronomy, Iowa State University, Ames, Iowa 50011, U.S.A.}

  \author{Sergey L. Bud'ko}
 \author{Paul C. Canfield}
\affiliation{Department of Physics and Astronomy, Iowa State University, Ames, Iowa 50011, U.S.A.}
\affiliation{Ames Laboratory, US DOE, Iowa State University, Ames, Iowa 50011, U.S.A}%
 
\date{\today}

\begin{abstract}
We report synthesis of hexagonal LaV$_x$Cr$_{1-x}$Ge$_3$ (\textit{x} = 0 -- 0.21, 1.00) single crystals, and present a systematic study of this series by measurements of temperature and field dependent magnetic susceptibility, magnetization, resistivity, and specific heat. Ferromagnetism has been observed for \textit{x} = 0 -- 0.21, and the system manifests a strong axial anisotropy in its ordered state. The decrease of the saturated moment and the effective moment per Cr with the increase of V-concentration suggests this is an itinerant ferromagnetic system. The Curie temperature declines monotonically as the V-concentration increases. Single crystalline samples could only be grown for $x$-values up to 0.21 for which the transition temperature was suppressed down to 36 K. Although we could not fully suppress $T_{\rm C}$ via V-substitution, for \textit{x} = 0.16, we performed magnetization measurements under pressure. The ferromagnetic state is suppressed under pressure at an initial rate of $dT_{\rm C}/dp$ $\simeq$ -- 11.7 K/GPa and vanishes by 3.3 GPa. 
\end{abstract}


\maketitle

\section{\label{sec:level1a}Introduction}

Transition metals and their compounds, can manifest itinerant magnetic behavior, with their magnetic properties originating from delocalized \textit{d}-electrons.\cite{Schlenker_2005, Uhlarz_2004, Thessieu_1995} Unlike the localized \textit{f}-electrons in the rare earth elements, \textit{d}-electrons' orbitals can be significantly altered by the formation of chemical bonds. 3$d$ electrons often become part of the conduction band, propagating in the materials, thus, their wavefunctions are very different from those of localized 4$f$ electrons. This gives rise to the relatively large exchange interactions between the \textit{d}-electrons. Based on the Stoner criterion,\cite{Stoner} at a critical value of the density of states (DOS) and on-site repulsion, \textit{d}-electrons can spontaneously split into spin-up and spin-down sub-bands, which leads to ferromagnetic ordering. Although the Stoner theory\cite{Stoner} provides the grounds for understanding the itinerant ferromagnetic state, there are still questions left to be answered about the role of spin fluctuations and the quantum criticality in the itinerant ferromagnetic systems.

Itinerant ferromagnets are of particular interest for studying the mechanism of magnetism and superconductivity near a quantum critical point (QCP). Unlike the classical phase transitions driven by temperature, a quantum phase transition (QPT) at zero temperature is driven by non-thermal parameters.\cite{Sachdev_1999} A QCP is thought to be a singularity in the ground state, at which point the characteristic energy scale of fluctuations above the ground state vanishes.\cite{Sachdev_1999} In itinerant ferromagnets, the temperature dependence of the magnetic properties has often been interpreted in terms of spin fluctuations.\cite{Lonzarich_1986, Lonzarich_1988, Hertz_1976} With the spin fluctuations, an ordered ground state can change into a non-ordered state by crossing a QCP. Non-Fermi liquid behaviors of the materials associated with a QCP can often be observed, such as the temperature divergences of the physical properties.\cite{Stewart_1984, Stewart_2001, Stewart_2006, Bud'ko_2004, Bud'ko_2005, Mun_2013} Moreover, superconductivity has been discovered in the vicinity of a QCP in weakly ferromagnetic systems, such as in the case of UGe$_2$\cite{Taufour_2010, Saxena_2000, Flouquet_2001} and UCoGe\cite{UCoGe}. On the boundary of a ferromagnetic state at low temperatures, a strong longitudinal magnetic susceptibility and magnetic interactions may lead to a superconducting state.\cite{Saxena_2000, Appel_1980, Varma_1986} The parallel-spin quasiparticles in the ferromagnetic system should form pairs in odd-parity orbitals, based on Pauli Exclusion Principle. Theories suggest this type of superconductivity should be spin-triplet and magnetically mediated.\cite{Saxena_2000, Appel_1980, Varma_1986} Thus, the suppression of ferromagnetism and the search for a QCP in the itinerant ferromagnetic systems may offer a better understanding of the magnetically mediated superconductivity and non-Fermi liquid behaviors. Chemical doping, pressure and magnetic field are often used to tune the magnetic orderings, and drive the criticality. For example, a QCP emerges in Zr$_{1-x}$Nb$_x$Zn$_2$ when the doping level reaches $x_{\rm c}$ = 0.083,\cite{Fisk_2006} and in CePd$_{1-x}$Ni$_x$ when the doping level is 0.95.\cite{Stewart_2001} YbAgGe\cite{Bud'ko_2004} , YbPtBi\cite{Mun_2013} and YbRh$_2$Si$_2$\cite{YbRh2Si2} can be driven to field induced QCPs associated with a non-Fermi-liquid behavior in the resistivity. In the case of MnSi\cite{Thessieu_1995} and UGe$_2$,\cite{Taufour_2010} itinerant-electron magnetism disappears at a first order transition and a QPT appears as pressure is applied.

LaCrGe$_3$ was reported to be metallic and to order ferromagnetically at 78 K.\cite{Mar_2007, Avdeev_2013} It forms in a hexagonal perovskite type (space group $P6_3/mmc$) structure. Previous work suggests it is an itinerant ferromagnet, with an estimated effective moment, 1.4 $\mu_{\rm B}$/f.u., significantly lower than the expected values of Cr$^{4+}$ (2.8 $\mu_{\rm B}$) or Cr$^{3+}$ (3.8 $\mu_{\rm B}$).\cite{Mar_2009} LaVGe$_3$, adopted the same crystal structure, is found to be non-magnetic above 2 K.\cite{Mar_2009} Based on the band structure calculated for both compounds, it is claimed that they have very similar DOS features and can probably be explained by the rigid band model.\cite{Mar_2007, Mar_2009} For LaCrGe$_3$, the \textit{d}-state of Cr manifests as a sharp peak near the Fermi level in the DOS, consistent with itinerant ferromagnetism as suggested by the Stoner model.\cite{Stoner} LaVGe$_3$, with fewer electrons, fills the band up to a lower energy level. Thus, the Fermi level of LaVGe$_3$ lies at a local minimum of the DOS, and shows paramagnetic behavior.

To suppress the ferromagnetism in this system, substituting V for Cr in LaV$_x$Cr$_{1-x}$Ge$_3$ is one of the rational choices, since this is expected to tune the DOS by changing the position of the Fermi level. Studies of polycrystalline samples show that V-substitution does change the magnetic exchange interactions, and the long-range magnetic ordering is suppressed.\cite{Mar_2009} Only the temperature dependence of magnetization was measured on the polycrystalline samples, and the precise stoichiometry of this doped system was not analyzed by chemical or physical measurement. The V-concentration dependence of Curie temperature was not reported, and it is not clear at which concentration the ferromagnetism is fully suppressed. Detailed measurements of transport and thermodynamic properties of the doped system are needed, since they allow to follow the evolution of the ferromagnetism and shed light on its mechanism.

Besides chemical substitution, an itinerant magnetic system can often be perturbed by applying pressure. Thus, for the LaV$_x$Cr$_{1-x}$Ge$_3$ series, pressure can also be used to suppress the magnetic state and discover a possible QCP.

In this work, we report the synthesis of single crystalline LaV$_x$Cr$_{1-x}$Ge$_3$ (\textit{x} = 0 -- 0.21, 1.00) samples, and present a systematic study of their transport and thermodynamic properties. A ferromagnetic transition has been confirmed, and the system shows a strong axial anisotropy in its ordered state. Both of the effective moment and the saturated moment per Cr decrease systematically as V-concentration increases, indicating that the Cr moment is fragile and the system is an itinerant ferromagnet. The Curie temperature decreases with V-substitution. The magnetic ordering is suppressed down to 36 K for the highest level of V-substitution obtained ($x$ = 0.21, other than the non-magnetic LaVGe$_3$). Given that ferromagnetism is not completely suppressed by our highest level of V-substitution, measurements of magnetization under pressure were performed on the $x$ = 0.16 sample. The ferromagnetic state is suppressed by the increasing pressure and vanishes by 3.3 GPa.

\section{\label{sec:level1b}Experimental Details}

\begin{figure}
\resizebox*{8.5cm}{!}{\includegraphics{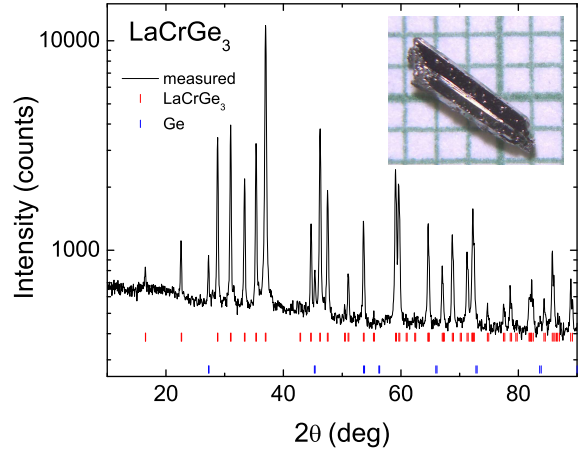}}
\caption{\label{fig:X-ray} (color online) Powder X-ray diffraction pattern of LaCrGe$_3$. Inset: Photo of a single crystalline LaCrGe$_3$ sample on a millimeter grid.}
\end{figure}

\begin{figure}
\resizebox*{8.5cm}{!}{\includegraphics{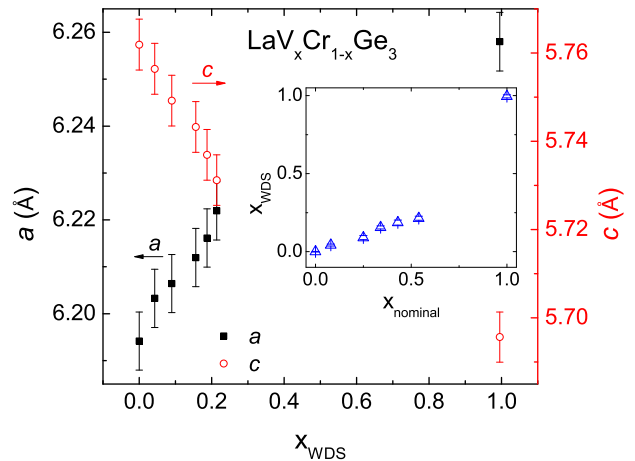}}
\caption{\label{fig:Lattice} (color online) The lattice parameters $a$ and $c$ of single crystalline LaV$_x$Cr$_{1-x}$Ge$_3$ compounds vs. V-concentration $x_{\rm WDS}$ measured by WDS. Inset: $x_{\rm WDS}$ vs. $x_{\rm nominal}$.}
\end{figure}

\begin{table*}
\caption{\label{tab:WDS}The WDS data (in atomic \%)for LaV$_x$Cr$_{1-x}$Ge$_3$. $N$ is the number of points measured on one sample, $x_{\rm nominal}$ is the nominal concentration, $x_{\rm WDS}$ is the average $x$ value measured, and 2$\sigma$ is two times the standard deviation of $x_{\rm WDS}$ from the $N$ values measured.}
\begin{ruledtabular}
\begin{tabular}{cccccccc}

$x_{\rm nominal}$	&	$N$	&	La	&	V	&	Cr	&	Ge	&	$x_{\rm WDS}$	&	2$\sigma$\\
\hline
0.00	&	13	&	20.06	&	0.01	&	19.80	&	60.13	&	0.00	&	0	\\
0.08	&	14	&	20.00	&	0.83	&	18.74	&	60.42	&	0.04	&	0.01	\\
0.25	&	12	&	19.98	&	1.76	&	17.89	&	60.36	&	0.09	&	0.01	\\
0.34	&	12	&	20.01	&	3.06	&	16.50	&	60.43	&	0.16	&	0.01	\\
0.43	&	16	&	20.04	&	3.69	&	16.09	&	60.19	&	0.19	&	0.02	\\
0.54	&	12	&	20.41	&	4.29	&	15.75	&	59.49	&	0.21	&	0.01	\\
1.00	&	14	&	19.66	&	20.46	&	0.09	&	59.79	&	1.00	&	0.01	\\

\end{tabular}
\end{ruledtabular}
\end{table*}

The relatively deep eutectic in the Cr-Ge binary \cite{Binary} provides an opportunity of growing LaV$_x$Cr$_{1-x}$Ge$_3$ compounds out of high-temperature solutions. \cite{Canfield_1992, Canfield_2010} Single crystals of LaV$_x$Cr$_{1-x}$Ge$_3$ were synthesized with a ratio of La:V:Cr:Ge = (13+2$x$):10$x$:(13--13$x$):(74+$x$) ($0\leq x \leq 0.6$). High purity ($>$ 3N) elements La, V, Cr and Ge were pre-mixed by arc-melting. The ingot was then loaded into a 2 ml alumina crucible and sealed in a fused silica tube under a partial pressure of high purity argon gas. The ampoule containing the growth materials was heated up to 1100 $^\circ$C over 3 h and held at 1100 $^\circ$C for another 3 h. The growth was then cooled to 825 $^\circ$C over 65 h at which temperature the excess liquid was decanted using a centrifuge.\cite{Canfield_1992, Canfield_2010} For $x$ = 1.0, i.e. LaVGe$_3$, excess Ge flux was used with an initial composition of La:V:Ge = 15:10:75, and the decanting temperature was adjusted to 880 $^\circ$C accordingly. Single crystals of LaV$_x$Cr$_{1-x}$Ge$_3$ grew as hexagonal rods with typical size of $\sim$ 0.7 $\times$ 0.7 $\times$ 5 mm$^3$ (seen in the inset of fig. \ref{fig:X-ray}). A considerable amount of second phase material was grown as the result of secondary solidification, which was identified to be V$_{11}$Ge$_8$ by powder X-ray diffraction. For growths with initial composition of $0.6 < x < 1.0$, the sizes of crystals dramatically decreased to submillimeters, and could not be visually distinguished from the secondary solidification (V$_{11}$Ge$_8$). Despite multiple attempts, single crystalline LaV$_x$Cr$_{1-x}$Ge$_3$ samples with higher $x$, which are distinguishable from the secondary solidification, could not be grown.  

Powder X-ray diffraction data were collected on a Rigaku MiniFlex II diffractometer with Cu K$\alpha$ radiation at room temperature. Samples with rod-like shape were selected for the measurement. Data collection was performed with a counting time of 2 s for every 0.02 degree. The Le Bail refinement was conducted using the program Rietica.\cite{X-ray} Error bars associated with the values of the lattice parameters were determined by statistical errors, and Si powder standard was used as an internal reference.

Elemental analysis of the samples was performed using wavelength-dispersive X-ray spectroscopy (WDS) in a JEOL JXA-8200 electron probe microanalyzer. Only clear and shiny, as grown surface regions were selected for determination of the sample stoichiometry, i.e. regions with residual Ge flux were avoided. For each compound, the WDS data were collected from multiple points on the same sample. 

Measurements of field and temperature dependent magnetization were performed in a Quantum Design, Magnetic Property Measurement System (MPMS). The ac resistivity was measured by a standard four-probe method in a Quantum Design, Physical Property Measurement System (PPMS). Platinum wires were attached to the sample using Epo-tek H20E silver epoxy, with the current flowing along the $c$-axis. The absolute values of resistivity are accurate to $\pm 15\%$ due to the accuracy of measurements of electrical contacts' positions.

Temperature dependent specific heat in zero field was measured in a PPMS using the relaxation technique for representative samples. The specific heat of LaVGe$_3$ was used to estimate the non-magnetic contributions to the specific heat of LaV$_x$Cr$_{1-x}$Ge$_3$. The magnetic contribution to specific heat from the Cr ions was calculated by the relation: $C_{\rm M}$ = $C_{\rm p}$(LaV$_x$Cr$_{1-x}$Ge$_3$) -- $C_{\rm p}$(LaVGe$_3$).

The temperature dependent, field-cooled magnetization of a single crystal for $x = 0.16$ under pressure was measured in a Quantum Design MPMS-SQUID magnetometer in a magnetic field of 20 Oe, 50 Oe and 1 kOe applied along the $c$-axis. Pressures of up to $4.9$ GPa were achieved with a moissanite anvil cell \cite{Pressure}. The body of the cell is made of Cu-Ti alloy and the gasket is made of Cu-Be. Daphne 7474 was used as a pressure transmitting medium, and the pressure was determined at $77$ K by the ruby fluorescence technique.

\section{\label{sec:level1c}Results and Analysis}
\subsection{\label{sec:level2a}Crystal Stoichiometry and Structure}

The stoichiometry of the LaV$_x$Cr$_{1-x}$Ge$_3$ samples was inferred by WDS analysis. Table \ref{tab:WDS} summarizes the normalized results showing the atomic percent of each element. The ratio of La:(V+Cr):Ge stays roughly as 1:1:3. The variation is induced by systematic error, and the counting statistics suggest there should be 2$\%$ or less relative error due to counting. As shown in the inset of fig. \ref{fig:Lattice}, the ratio of $x_{\rm WDS}$ over $x_{\rm nominal}$ is approximately 0.4, and the small 2$\sigma$-values suggest that the samples are homogeneous. In the following, the measured, $x_{\rm WDS}$, rather than nominal $x$ values will be used to index the stoichiometry of the compounds in this series.

Powder X-ray diffraction patterns were collected on ground single crystals from each compound. Fig. \ref{fig:X-ray} presents the LaCrGe$_3$ X-ray pattern as an example. The main phase was refined with the known $P6_3/mmc$ (No. 194) structure. Small traces of Ge residue can be detected, whereas no clear evidence of La-Ge, V-Ge, or Cr-Ge binaries was found. Similar results ($P6_3/mmc$ structure with minority phase of Ge) were obtained for the rest of the series. The lattice parameters, obtained by the analysis of the powder X-ray diffraction data, are presented in fig. \ref{fig:Lattice}. As is shown, $a$ and $c$ are monotonically changing as the $x$ increases, which is consistent with the reported data.\cite{Mar_2009} Crystallographically, transition metal elements in LaCrGe$_3$ and LaVGe$_3$ occupy the same unique site $2a$.\cite{Mar_2007, Mar_2009} 

\subsection{\label{sec:level2b}Effects of chemical substitution on the physical properties of LaV$_x$Cr$_{1-x}$Ge$_3$}

\begin{figure}
\resizebox*{8.5cm}{!}{\includegraphics{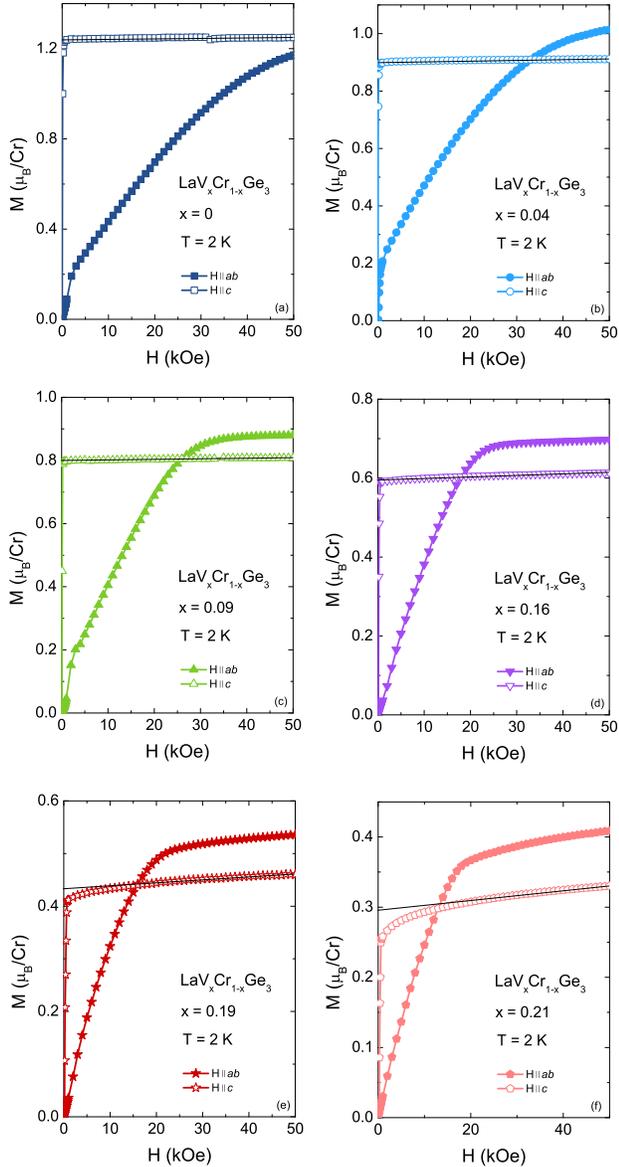}}
\caption{\label{fig:MH} (color online) Anisotropic field-dependent magnetization data for LaV$_x$Cr$_{1-x}$Ge$_3$ ($x$ = 0 -- 0.21) taken at 2 K. Fine solid lines through the high field \textbf{H} $\parallel c$ data extrapolate back to $H = 0$, $\mu_{\rm S}$ values shown in Table \ref{tab:Temperature}.}
\end{figure}

\begin{figure}
\resizebox*{8.5cm}{!}{\includegraphics{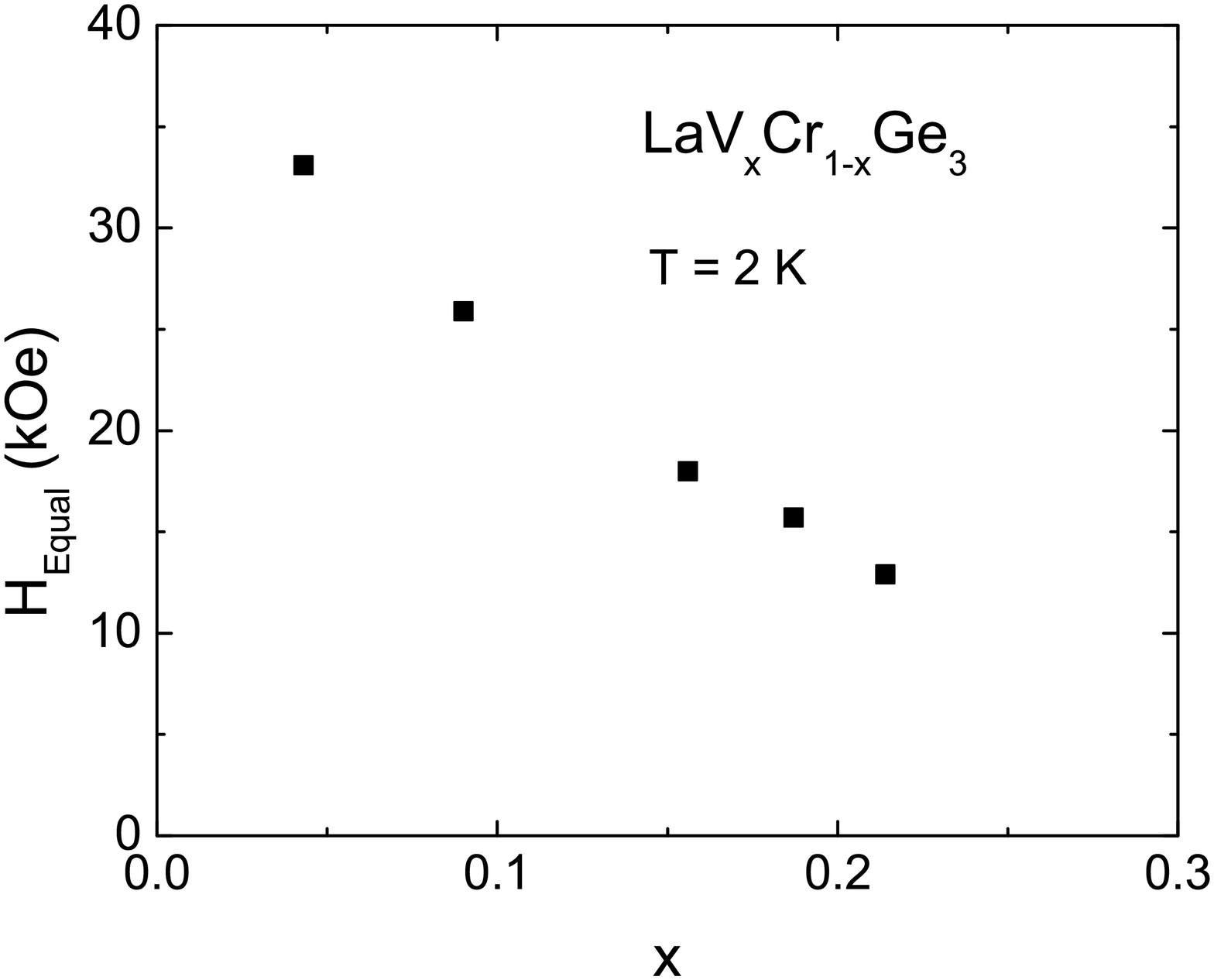}}
\caption{\label{fig:Equal} $H_{\rm Equal}$ (the field at which $M_{c}$ = $M_{ab}$) as a function of $x$ for LaV$_x$Cr$_{1-x}$Ge$_3$ ($x$ = 0.04 -- 0.21) taken at 2 K. }
\end{figure}

Figure \ref{fig:MH} (a)--(f) present the anisotropic field-dependent magnetization isotherms for the LaV$_x$Cr$_{1-x}$Ge$_3$ ($x$ = 0 -- 0.21) series. The measurements were performed with \textbf{H} parallel to $ab$-plane and $c$-axis at 2 K. For \textbf{H} $\parallel c$, the magnetization of all compounds saturates very rapidly as the magnetic field increases form $H$ = 0, a manifestation of a typical ferromagnetic behavior. This clearly shows that $c$-axis is the easy axis, consistent with the neutron powder diffraction study.\cite{Avdeev_2013} For \textbf{H} $\parallel ab$, the magnetization rises more slowly as the applied field increases. As can be seen, at low fields, $M_c \gg M_{ab}$; however, as $x$ increases, $H_{\rm Equal}$, the field at which $M_{c}$ equals $M_{ab}$ decreases monotonically (fig. \ref{fig:Equal}). The $x$-dependence of $H_{\rm Equal}$ presents a calliper of the diminishing range of the low-field $M_c > M_{ab}$ anisotropy. These data suggest that the LaV$_x$Cr$_{1-x}$Ge$_3$ compounds might have a canted ferromagnetic structure. The change of anisotropy is probably caused by field induced spin reorientation, which is consistent with previous neutron study.\cite{Avdeev_2013} For \textbf{H} $\parallel c$, the saturated moment $\mu_{\rm S}$ per Cr is determined by linear extrapolations of the magnetization from high fields to $H$ = 0. For $x$ = 0, $\mu_{\rm S}$ is found to be about 1.25 $\mu_{\rm B}$/Cr, essentially identical to the reported value 1.22 $\mu_{\rm B}$.\cite{Avdeev_2013} It monotonically decreases as V-concentration increases and drops to 0.30 $\mu_{\rm B}$/Cr for $x$ = 0.21. The values of the saturated moment $\mu_{\rm S}$  with \textbf{H} $\parallel c$ are summarized in Table \ref{tab:Temperature}. Again, the decrease of the saturated moment implies the LaV$_x$Cr$_{1-x}$Ge$_3$ series probably is an itinerant ferromagnetic system.\cite{Mar_2007} 

\begin{figure}
\resizebox*{8.5cm}{!}{\includegraphics{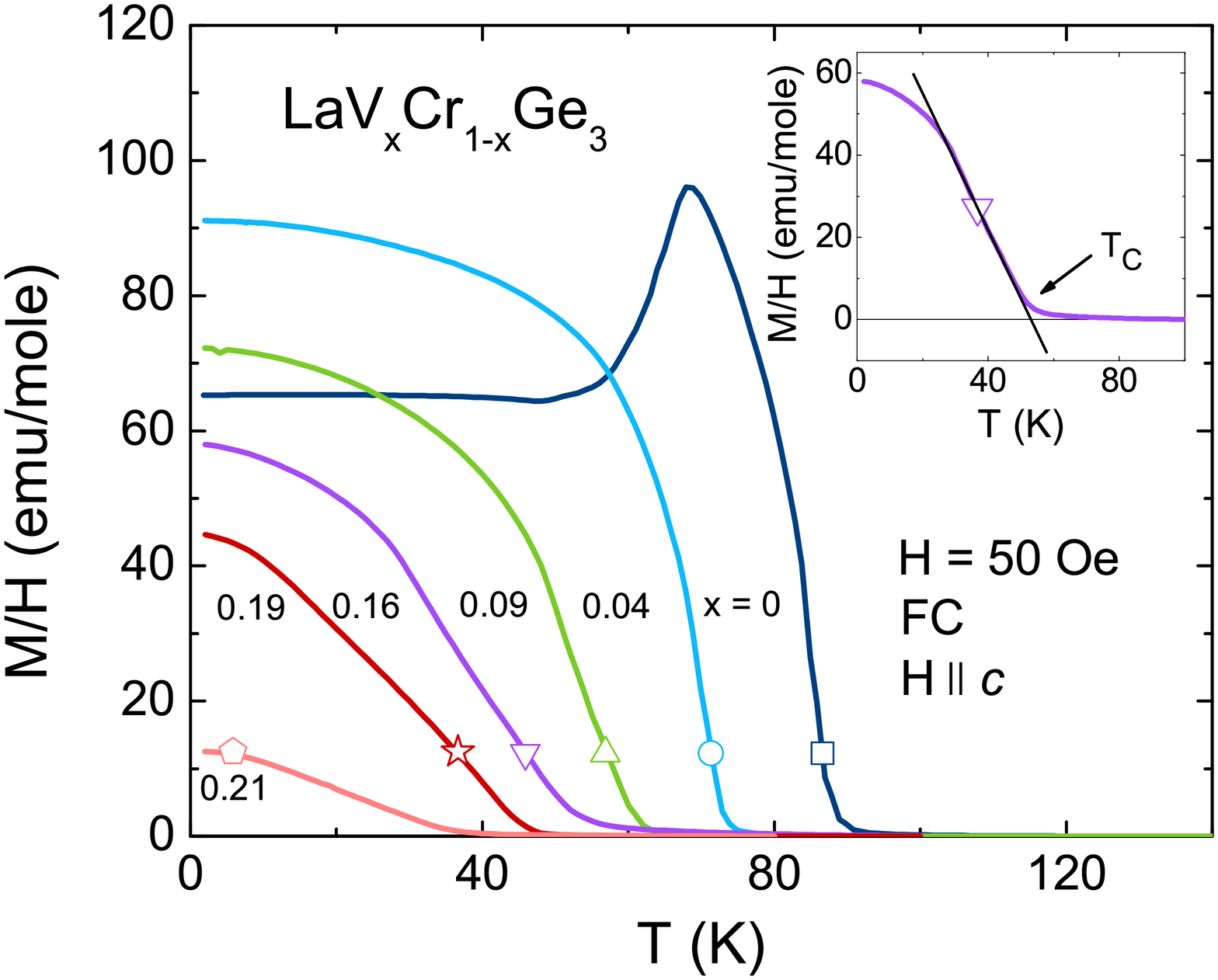}}
\caption{\label{fig:MT50} (color online) Field-cooled (FC) magnetization as a function of temperature for LaV$_x$Cr$_{1-x}$Ge$_3$ ($x$ = 0 -- 0.21) at 50 Oe with \textbf{H} $\parallel c$. Inset: Enlarged temperature dependence of magnetization near phase transition for $x$ = 0.16. The arrow shows the criterion used to infer the transition temperature.}
\end{figure}

To estimate the Curie temperature $T_{\rm C}$ from the magnetization measurements, we studied the temperature-dependent, field-cooled (FC) magnetization of the LaV$_x$Cr$_{1-x}$Ge$_3$ series, with \textbf{H} $\parallel c$ at 50 Oe, as shown in fig. \ref{fig:MT50}. The magnetization for LaCrGe$_3$ exhibits a sudden increase near 90 K, indicating a transition to a ferromagnetic state. However, at around 68 K, its value starts declining, then saturates at low temperatures, leaving a peak seen in its magnetization. This is probably associated with the changes of the magnetic domains and the demagnetization field upon cooling. Similar feature was not observed for the V-doped compounds. It is possibly due to the pinning effect brought by the V-substitution. For the other members of the series, the susceptibility shows the expected, rapid increase and the tendency to saturation at low temperatures, which indicate the existence of a ferromagnetic state in this series for $x$ up to 0.21. The Curie temperature was estimated by extrapolating the maximum slope in $M/H$ to zero, as shown by the arrow in the inset of fig. \ref{fig:MT50}; the $T_{\rm C}$ values are listed in Table \ref{tab:Temperature}. Given that these are very low field $M (T)$ data, these values should not be too different from those inferred from the Arrott plots, fig. \ref{fig:Arrott} (see below). The monotonic change of the Curie temperature demonstrates that the ferromagnetism in the LaV$_x$Cr$_{1-x}$Ge$_3$ series is systematically suppressed by the V-substitution. 

\begin{figure}
\resizebox*{8.5cm}{!}{\includegraphics{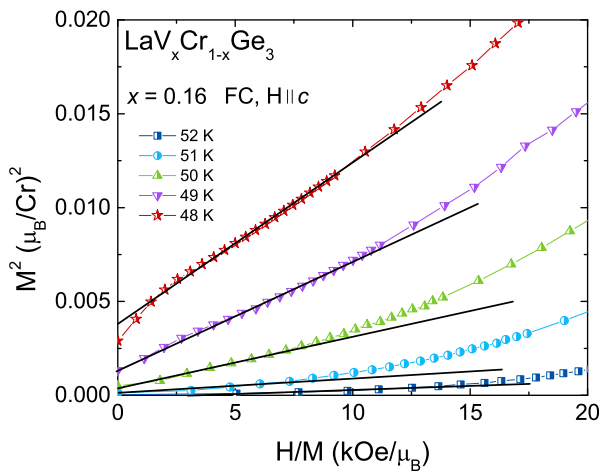}}
\caption{\label{fig:Arrott} (color online) The Arrott plot in the form of $M^2$ vs $H/M$ for $x$ = 0.16, with \textbf{H} $\parallel c$.}
\end{figure}

Given that a ferromagnet possesses a spontaneous magnetization below its Curie temperature, even without external magnetic field applied, the determination of the Curie temperature from the temperature-dependent magnetization is not without ambiguity. To have a better evaluation of the Curie temperature, magnetization isotherms in the vicinity of $T_{\rm C}$ were measured for $x$ = 0.16. Since the easy axis is parallel to the sample's long axis, and in order to reduce the uncertainty caused by demagnetization effects, magnetic field was applied along the rod (\textbf{H} $\parallel c$). According to the Arrott-Noakes\cite{Arrott, Noakes} relation: $(H/M)\propto M^2$ at $T_{\rm C}$, the ferromagnetic ordering temperature can be inferred from the magnetization data by noting the temperature at which the low-field data pass though the origin. As shown in fig. \ref{fig:Arrott}, the Curie temperature for $x$ = 0.16, determined by Arrott plot\cite{Arrott}, is about 51 $\pm$ 1 K, which is very close to the value obtained from the low field magnetization measurement (also seen in Table \ref{tab:Temperature}). Therefore, the $T_{\rm C}$ determined from the low field magnetization data appears to be reliable for these materials. It should be noted that the isothermal curves in the Arrott plot are nonlinear. Such feature may be associated with complex magnetic phenomena in a disordered system\cite{Williams_1986, Jia_2008} rather than with a simple clearly defined Landau type second order phase transition.  

\begin{table*}
\caption{\label{tab:Temperature} Summarized $\mu_{\rm S}$ (\textbf{H} $\parallel c$), $\mu_{\rm eff}$, $\theta_{\rm c}$ and ordering temperatures from magnetization $T_{\rm C}^{mag}$, resistivity $T_{\rm C}^{\rho}$, specific heat $T_{\rm C}^{C_{\rm p}}$, Arrott plot data and residual resistivity $\rho_0$ for LaV$_x$Cr$_{1-x}$Ge$_3$ ($x$ = 0 -- 0.21).}
\begin{ruledtabular}
\begin{tabular}{ccccccccc}
$x_{\rm WDS}$	&$\mu_{\rm S}$ ($\mu_{\rm B}$/Cr), \textbf{H} $\parallel c$	&$\mu_{\rm eff}$ ($\mu_{\rm B}$/Cr)	&	$\theta_{\rm c}$ (K)	&	$T_{\rm C}^{mag}$ (K)	&	$T_{\rm C}^{\rho}$ (K)	&	$T_{\rm C}^{C_{\rm p}}$ (K)&	$T_{\rm C}$(Arrott) (K)&$\rho_0$ ($\mu\Omega$ cm)	\\
\hline
0.00	&	1.25& 2.5	&	91.7	&	88	&	84	&	85&	&36\\
0.04	&	0.91& 2.3	&	84.7	&	73	&	68	&	69&	&78\\
0.09	&	0.81& 2.2	&	64.6	&	61	&	54	&	62& &116\\
0.16	&	0.59& 2.1	&	45.9	&	52	&	37	&	46&   51 $\pm$ 1&100	\\
0.19	&	0.43& 2.0	&	26.2	&	46	&	 	&	&	&97\\
0.21	&   0.30& 1.9	&	6.7	    &	36  &	 	&	&	&66\\
1.00	&     &  	&	   &	   &	 	&	&	&8\\
\end{tabular}
\end{ruledtabular}
\end{table*}

\begin{figure}
\resizebox*{8.5cm}{!}{\includegraphics{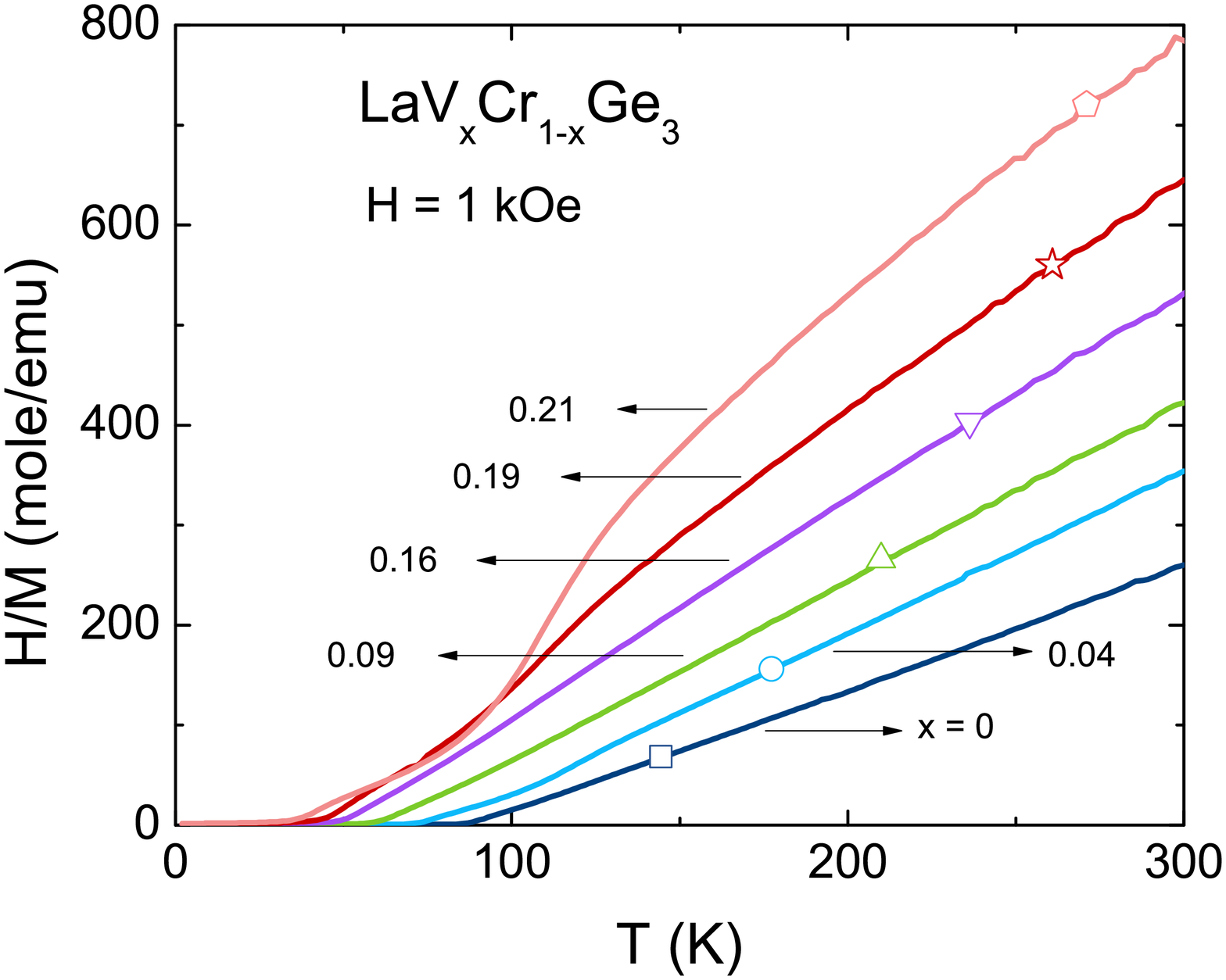}}
\caption{\label{fig:Eff} (color online) Temperature-dependent inverse susceptibility $H/M$ data for LaV$_x$Cr$_{1-x}$Ge$_3$ ($x$ = 0 -- 0.21) with $H$ = 1 kOe.}
\end{figure}

The inverse of polycrystalline average susceptibility $H/M$ measured at 1 kOe is shown in fig. \ref{fig:Eff}. The polycrystalline average susceptibility was estimated by $\chi_{ave}$ = $\frac{1}{3}$ $(\chi_{c}+2\chi_{ab})$. At high temperatures, all of the compounds follow the Curie-Weiss behavior. However, evident deviations from the Curie-Weiss law can be observed below 130 K for $x$ = 0.19 and 0.21, as also seen in GdFeZn$_{20}$.\cite{Jia_2008} Further investigations are needed to understand the origin of these deviations. The temperature range of 150 K to 300 K was selected for fitting the high-temperature magnetic susceptibility with $1/\chi=(T-\theta_{\rm c})/C$, where $\theta_{\rm c}$ is the Curie-Weiss temperature and $C$ is the Curie constant. The effective moments $\mu_{\rm eff}$ and $\theta_{\rm c}$ are summarized in Table \ref{tab:Temperature}. Considering the presence of small amount of Ge and V$_{11}$Ge$_8$ as well as the accuracy of measuring sample's mass, the values of $\mu_{\rm eff}$ in this series are accurate to $\pm 5\%$. For $x$ = 0, $\mu_{\rm eff}$ is found to be 2.5 $\mu_{\rm B}$/Cr, a value that is smaller than the calculated value for Cr$^{4+}$: 2.8 $\mu_{\rm B}$ or Cr$^{3+}$: 3.8 $\mu_{\rm B}$, yet larger than the reported value: 1.4 $\mu_{\rm B}$.\cite{Mar_2007} The effective moment per Cr ion decreases as the V-concentration increases. Also shown in fig. \ref{fig:Eff}, as $x$ increases, the slope of the $H/M$ curve rises gradually, and the Curie-Weiss temperature decreases from 91.7 K for $x$ = 0 to 6.7 K for $x$ = 0.21 monotonically. The positive $\theta_{\rm c}$ values indicate that ferromagnetic interactions are dominant in this series. The decrease in $\theta_{\rm c}$ suggests that the ferromagnetic interaction is suppressed by V doping. Based on all of these results, it is highly likely that the Cr ions in the LaV$_x$Cr$_{1-x}$Ge$_3$ compounds manifest non-local-moment like behavior.  

\begin{figure}
\resizebox*{8.5cm}{!}{\includegraphics{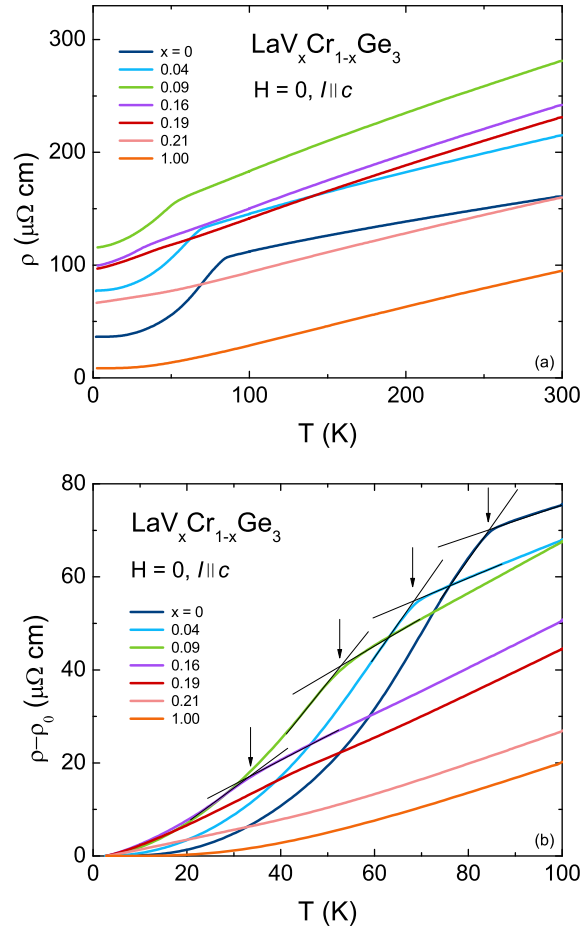}}
\caption{\label{fig:RT} (color online) (a) The temperature dependent-electrical resistivity for the LaV$_x$Cr$_{1-x}$Ge$_3$ compounds. (b) Enlarged temperature-dependent $\rho$ -$\rho_0$ at low temperatures. The transition temperatures are shown by the arrows.}
\end{figure}

The electrical resistivity as a function of temperature for LaV$_x$Cr$_{1-x}$Ge$_3$ is presented in fig. \ref{fig:RT} (a). At high temperatures, the electrical resistivity drops linearly upon cooling, characteristic of normal metallic behavior. For LaCrGe$_3$, due to the loss of spin disorder scattering, a clear break in resistivity occurs at about 84 K. With the subtraction of the residual resistivity $\rho_0$ (listed in Table \ref{tab:Temperature}), the evolution of the ferromagnetic transition with increasing $x$ can be clearly seen in fig. \ref{fig:RT} (b). As the V-doping level increases, the spin disorder scattering associated with the Cr moment ordering becomes broadened . For $x$ = 0.19 and 0.21, the feature is too subtle to be clearly detected. Due to the broadening transition feature, determining $T_{\rm C}$ via $d\rho/dT$ is problematic. Instead, the point at which the slope of $\rho (T)$ changes is used to infer the critical temperature in the resistivity data, as indicated by the arrows in fig. \ref{fig:RT} (b). The inferred $T_{\rm C}^{\rho}$ is summarized in Table \ref{tab:Temperature}, and it is clear that the Curie temperature decreases monotonically as the V-concentration increases. In addition, $\rho_0$ seems showing a broad maximum as $x$ increases, which is likely due to more disorder/impurities induced by substitution.

\begin{figure}
\resizebox*{8.5cm}{!}{\includegraphics{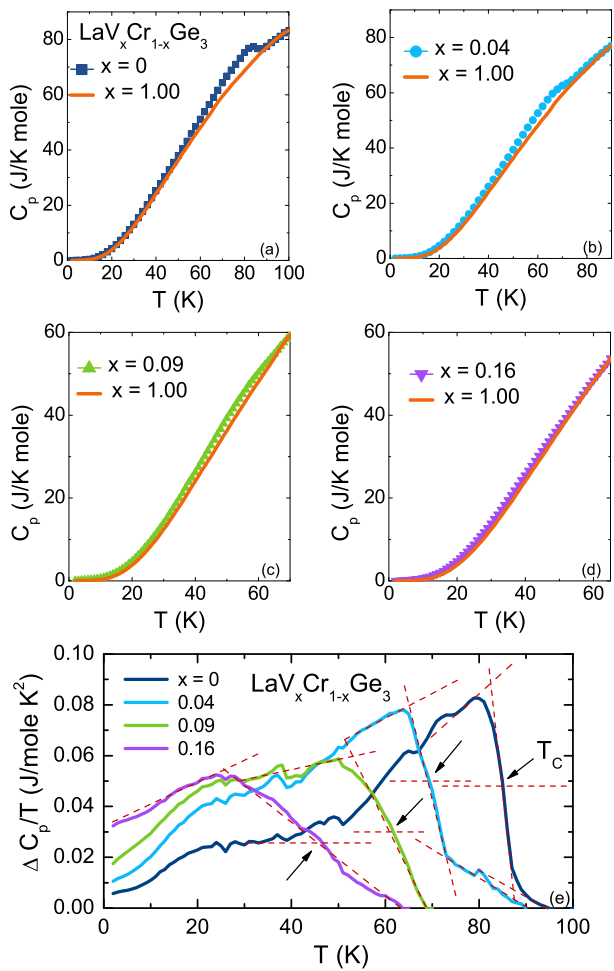}}
\caption{\label{fig:CT} (color online) Temperature-dependent of specific heat for LaV$_x$Cr$_{1-x}$Ge$_3$ with (a) $x$ = 0 and 1.00, (b) $x$ = 0.04 and 1.00, (c) $x$ = 0.09 and 1.00, (d) $x$ = 0.19 and 1.00. (e) Magnetic contributions to the specific heat as a function of temperature for LaV$_x$Cr$_{1-x}$Ge$_3$ ($x$ = 0 -- 0.16). The arrows show the criteria used to infer the transition temperature.}
\end{figure}

The temperature-dependent specific heat data for the LaV$_x$Cr$_{1-x}$Ge$_3$ ($x$ = 0, 0.04, 0.09, 0.16 and 1.00) series are presented in fig. \ref{fig:CT} (a) -- (d). The specific heat can be estimated by the relation $C_{\rm p}(T)$ = $C_{\rm e}$ + $C_{\rm ph}$ + $C_{\rm M}$, where $C_{\rm e}$ is the conduction electron contribution, $C_{\rm ph}$ is the phonon contribution, and $C_{\rm M}$ is the magnetic contribution. Since LaVGe$_3$ is non-magnetic, $C_{\rm e}$ + $C_{\rm ph}$ can be approximated by the $C_{\rm p}$ data of LaVGe$_{3}$. Thus, the magnetic contribution $C_{\rm M}$ can be evaluated by the relation: $C_{\rm M}$ = $C_{\rm p}$(LaV$_x$Cr$_{1-x}$Ge$_3$) -- $C_{\rm p}$(LaVGe$_3$). Above the ordering temperature, the specific heat of all compounds are expected to behave in the same manner. Therefore, $C_{\rm p}(T)$ of all compounds were normalized with respect to LaVGe$_{3}$'s specific heat $C_{\rm p}$(LaVGe$_{3}$), with the highest temperature $C_{\rm p}$ values setting to be equal (as seen in fig. \ref{fig:CT} (a) -- (d)). The changes induced by the normalization are less than 3$\%$. An anomaly can be observed in the $C_{\rm p}$(LaV$_x$Cr$_{1-x}$Ge$_3$) with the comparison of $C_{\rm p}$(LaVGe$_{3}$). This anomaly, associated with the ferromagnetic transition, can be best seen in LaCrGe$_{3}$ sample, at $\sim$ 85 K. As V-doping level increases, the feature becomes less obvious and systematically shifts to lower temperatures. For $x\geq$ 0.19, this feature is no longer detectable. To estimate the ordering temperature, $\bigtriangleup C_p/T$ for $x$ = 0, 0.04, 0.09, 0.16 are plotted in fig. \ref{fig:CT} (e). The magnetic phase transition manifests itself as a local maximum. The change of slope seen at $\sim$ 87 K for $x$ = 0 and $\sim$ 73 K for $x$ = 0.04 may indicate the on-set of the transition. The mid-point on the rise of $\bigtriangleup C_p/T$ was chosen as the criteria for $T_{\rm C}^{C_{\rm p}}$, as indicated by the arrows in the plot. These $T_{\rm C}^{C_{\rm p}}$ values are also presented in Table \ref{tab:Temperature}. Again, we observe that with the increasing amount of V substituted for Cr, the ferromagnetic state in this series is gradually suppressed. The magnetic entropy per mole Cr ($S_{\rm M}$) is estimated by the integration of its magnetic contribution to the specific heat. For $x$ = 0, $S_{\rm M}$ is found to be about 56$\%$ of $R$ln2 at $T_{\rm C}$, which again suggests that Cr ions show non-local-moment like behavior. For the other members in this series, due to the broadening transition feature, the estimation of $S_{\rm M}$ become less precise.

\subsection{\label{sec:level2c}Effects of pressure on the magnetic properties of LaV$_x$Cr$_{1-x}$Ge$_3$}

\begin{figure*}
\resizebox*{16cm}{!}{\includegraphics{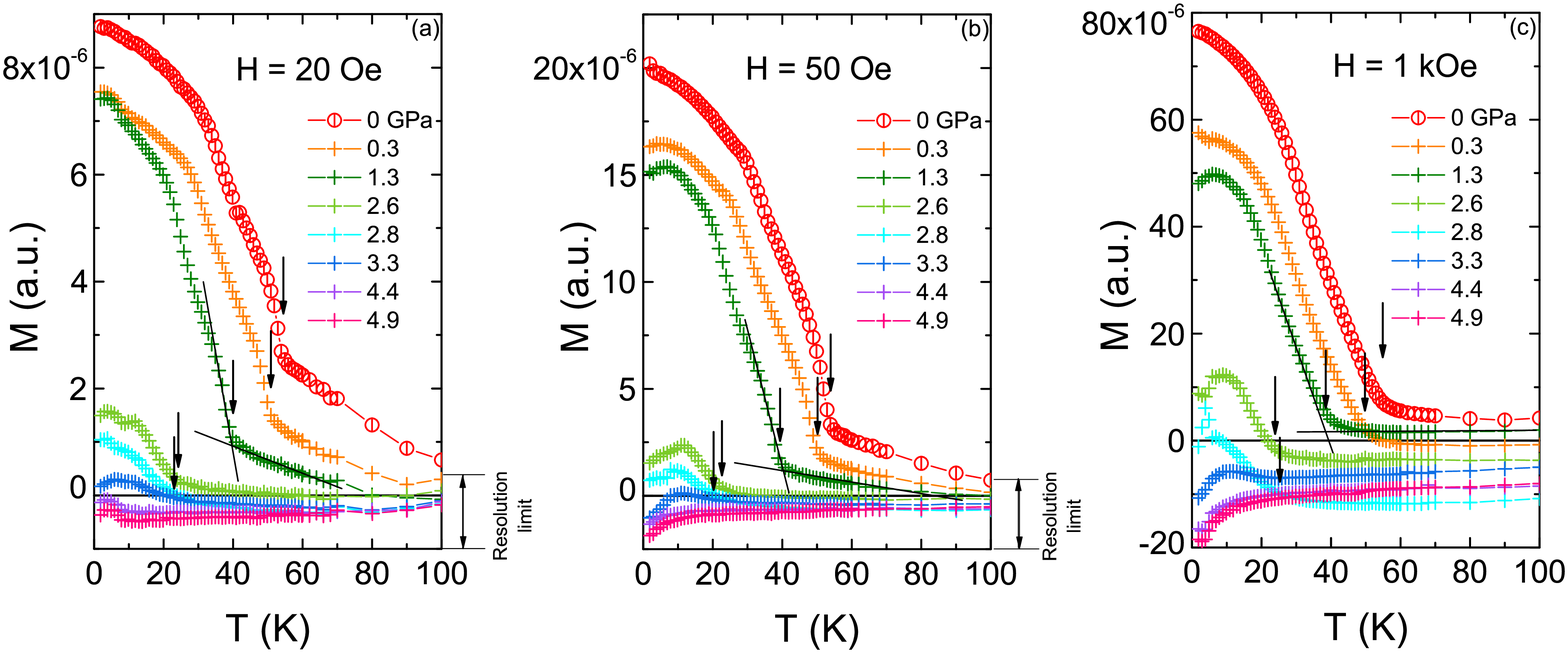}}
\caption{\label{fig:Pressure} (color online) Temperature dependence of the field-cooled magnetization for $x$ = 0.16 under different pressures with \textbf{H} $\parallel c$ at (a) 20 Oe, (b) 50 Oe and (c) 1 kOe. Arrows indicate the criteria for the determination of the Curie temperature $T_{\rm C}$.}
\end{figure*}

\begin{figure}
\resizebox*{8.5cm}{!}{\includegraphics{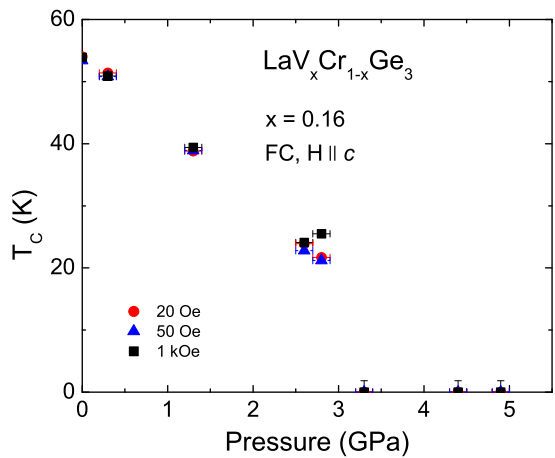}}
\caption{\label{fig:P_PD} (color online) Pressure dependence of $T_{\rm C}$ for $x$ = 0.16 measured at 20 Oe, 50 Oe and 1 kOe.}
\end{figure}

Given that i) we could only grow single crystals for $x \leq$ 0.21, and ii) that up to $x$ = 0.16 the ferromagnetic transition can be confirmed in different measurements, we decided to evaluate the potential for quantum critical behavior by using pressure as a second tuning parameter. Figure \ref{fig:Pressure} (a) -- (c) show the temperature dependence of the field-cooled magnetization for $x$ = 0.16 measured under different pressures. The measurements were performed with \textbf{H} $\parallel c$ and $H$ = 20 Oe, 50 Oe and 1 kOe. The Curie temperature $T_{\rm C}$ is revealed by a rather sharp increase of the magnetization. Due to the loss of the signal, there is a serious limitation to the determination of $T_{\rm C}$ close to the critical pressure. For higher field, $H$ = 1 kOe (fig. \ref{fig:Pressure} (c)), measurements and data analysis are limited to the large background of the pressure cell (This is the most likely source of apparent diamagnetic shifts in higher pressure data). By comparing fig.\ref{fig:Pressure} (a), (b) and (c), the magnetization under 3.3 GPa is not considered as a ferromagnetic behavior. The pressure dependences of the Curie temperature measured at different fields show consistent behaviors, as plotted in fig.\ref{fig:P_PD}. The result shows $T_{\rm C}$ decreases with applied pressure at an initial rate of $dT_{\rm C}/dp$ $\simeq$ -- 11.7 K/GPa below 2.8 GPa, and no ferromagnetic transition can be detected in our measurements above 3.3 GPa. Similarly, the low temperature magnetization decreases as $T_{\rm C}$ decreases with applied pressure as shown in fig. \ref{fig:Pressure} (a) -- (c). Although the low temperature signal is not necessarily equal to the saturation magnetization, the decrease of the low temperature magnetization following the decrease of $T_{\rm C}$ is expected for an itinerant ferromagnet \cite{Moriya_1973} and was experimentally observed in ZrZn$_2$.\cite{Uhlarz_2004, Huber_1975}

\section{\label{sec:level1d}Discussion and Conclusions }

\begin{figure}
\resizebox*{8.5cm}{!}{\includegraphics{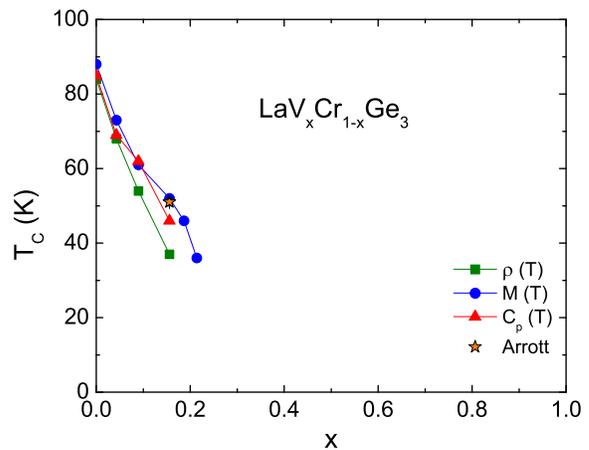}}
\caption{\label{fig:PD} (color online) $x$-dependent $T_{\rm C}$ for LaV$_x$Cr$_{1-x}$Ge$_3$ determined by $M(T)$, $\rho(T)$ and $C_{\rm p}(T)$ measurements as well as Arrott plot.}
\end{figure}

\begin{figure}
\resizebox*{8.5cm}{!}{\includegraphics{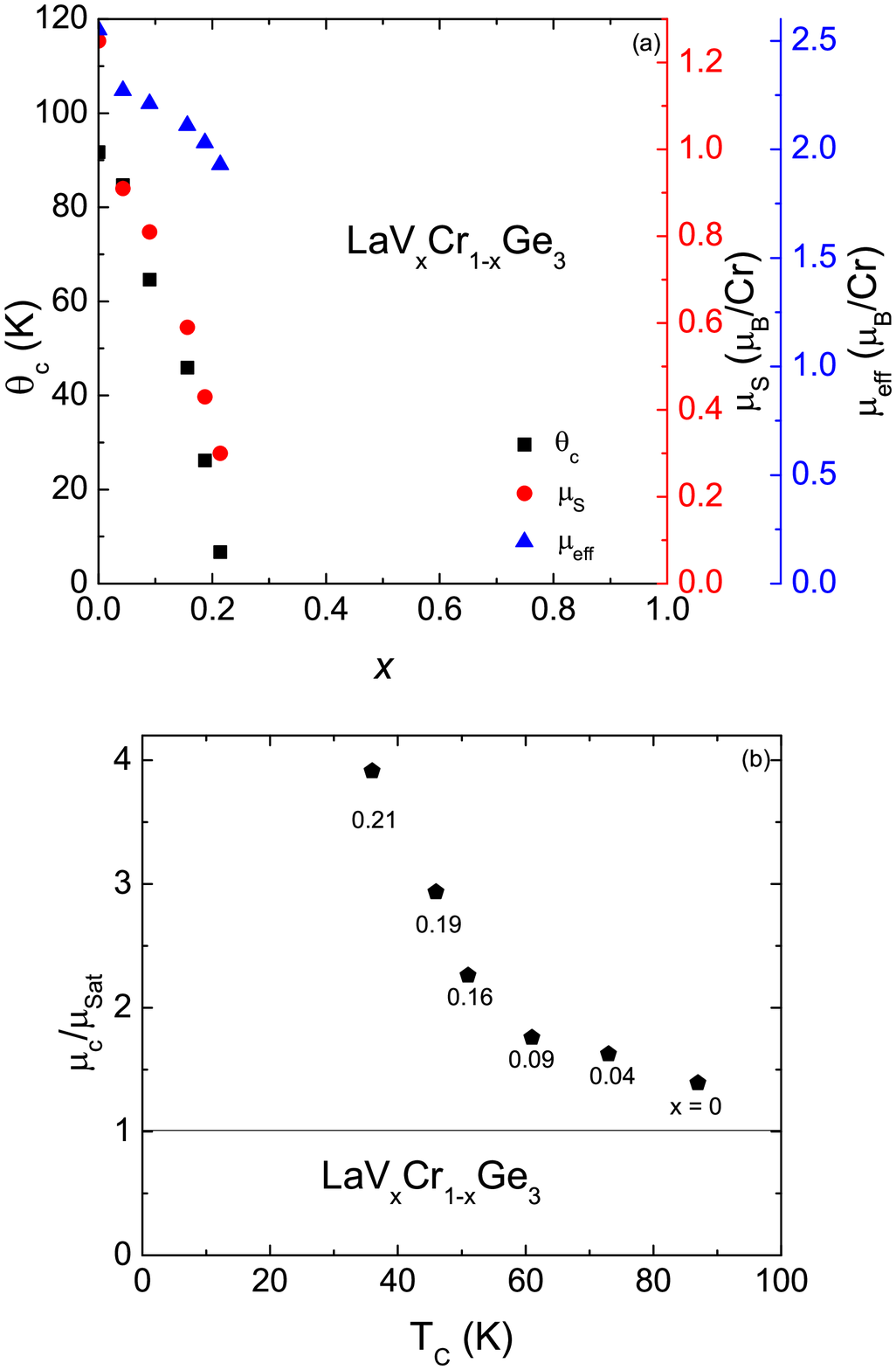}}
\caption{\label{fig:Moments} (color online) (a) The Curie-Weiss temperature $\theta_{\rm c}$, saturated moment $\mu_{\rm S}$ along $c$-axis and effective moment $\mu_{\rm eff}$ per Cr as a function of $x$ for LaV$_x$Cr$_{1-x}$Ge$_3$. (b) The Rhodes-Wohlfarth ratio $\mu_{\rm c}$/$\mu_{\rm S}$ as a function of Curie temperature $T_{\rm C}$.}
\end{figure}

The growth of single crystalline LaV$_x$Cr$_{1-x}$Ge$_3$ (\textit{x} = 0 -- 0.21, 1.00) samples has allowed for the detailed study of the anisotropic properties, the determination of the easy axis as well as the estimate of the effective moment and the saturated moment. In addition, careful chemical analysis was performed to determine the precise concentration of this doped system. This offers a clearer understanding of chemical substitution effect on the suppression of the ferromagnetism in this system, and is also crucial for calculating the saturated and effective moment per Cr ion.

We have been able to suppress the ferromagnetism in the LaV$_x$Cr$_{1-x}$Ge$_3$ series via chemical substitution. The ordering temperatures inferred from low field magnetization, resistivity and specific heat measurements are summarized in Table \ref{tab:Temperature}. A phase diagram of $x$-dependent $T_{\rm C}$ for LaV$_x$Cr$_{1-x}$Ge$_3$ was assembled in fig. \ref{fig:PD}. For $x$ = 0.19 and 0.21, magnetic transitions can only be detected in $M(T)$ but not in $\rho(T)$ and $C_{\rm p}(T)$ measurements. We can see that for the LaV$_x$Cr$_{1-x}$Ge$_3$ series, the ferromagnetic transition temperature is suppressed almost linearly by V doping: $T_{\rm C}$ = 88 K for $x$ = 0, and $T_{\rm C}$ = 36 K for $x$ = 0.21. Since single crystalline LaV$_x$Cr$_{1-x}$Ge$_3$ compounds with 0.21 $< x <$ 1.00 were not synthesized, the exact concentration $x_{\rm c}$ at which the ferromagnetism in this series is completely suppressed via V substitution is not determined. Based on the our data, a critical concentration is likely to exist near $x$ = 0.3. It is worth noting that this is the substitution range that a linear extrapolation of the $H_{\rm Equal}$ data shown in fig. \ref{fig:Equal} reaches zero.

The estimated $\mu_{\rm S}$ and $\mu_{\rm eff}$ per Cr as a function of $x$ are plotted in fig. \ref{fig:Moments} (a). As is shown, both $\mu_{\rm S}$ and $\mu_{\rm eff}$ decrease in a clear manner as the V-concentration increases. Consistent with the Stoner model, this suggests the system possesses a fragile ferromagnetism which can be easily perturbed. The criterion for the ferromagnetic state is given by the relation $U D(\varepsilon_F) \geq 1$, where $U$ and $D(\varepsilon_F)$ are Coulomb repulsion and the DOS at the Fermi level, respectively.\cite{Stoner} Given the fact that $T_{\rm C}$ decreases as $x$ increases, it is likely that $U$ and/or $D(\varepsilon_F)$ is changed by V-substitution in the LaV$_x$Cr$_{1-x}$Ge$_3$ system. With the increasing level of V-doping, the ferromagnetism is continuously suppressed, and will eventually disappear at a critical V-concentration $x_{\rm c}$. However, due to the lack of higher V-doped samples, $x_{\rm c}$ can not be identified precisely in this study. Similarly, in the case of Curie-Weiss temperature, clear suppression in $\theta_{\rm c}$ by V-doping can be observed, as shown in fig. \ref{fig:Moments} (a). Again, this implies the ferromagnetic interaction is weakened by V-substitution. Given the values of $\mu_{\rm S}$ and $\mu_{\rm eff}$, the Rhodes-Wolfarth ratio (RWR)\cite{RWR} can be calculated, seen in fig. \ref{fig:Moments} (b). According to Rhodes and Wolfarth, RWR = $\mu_{\rm c}$/$\mu_{\rm S}$, where $\mu_{\rm c}$ is related to the number of moment carriers, and can be obtained from $\mu_{\rm c}$($\mu_{\rm c}$+1)=$\mu_{\rm eff}^2$. While RWR = 1 is an indication of localized magnetism, larger RWR values suggest the existence of itinerant ferromagnetism. In our case, RWR of LaV$_x$Cr$_{1-x}$Ge$_3$ ranges from $\simeq$ 1.4 for $x$ = 0 to $\simeq$ 3.9 for $x$ = 0.21, suggesting the ferromagnetism is itinerant. In addition, as V-concentration increases, the change of RWR as a function of $T_{\rm C}$ exhibits very similar behavior as seen in the original Rhodes-Wohlfarth plot.\cite{RWR} It should be noted that the suppression of ferromagnetism does not necessarily lead to a QPT, new magnetic state, such as spin glass, may also emerge.\cite{Schlenker_2005, Mydosh} However, in the case of the LaV$_x$Cr$_{1-x}$Ge$_3$ series, given the RWR ratio and the fact that both $\mu_{\rm S}$ and $\mu_{\rm eff}$ decrease as the V-concentration increases, it is promising for it being a potential QCP system. \cite{Fisk_2006, RWR} 

We further suppressed the ferromagnetism for $x$ = 0.16 by pressure up to 4.9 GPa. As seen in fig. \ref{fig:P_PD}, the Curie temperature decreases as the applied pressure increases, at an initial rate of $dT_{\rm C}/dp$ $\simeq$ -- 11.7 K/GPa below 2.8 GPa. The ferromagnetic signal vanishes at $\simeq$ 3.3 GPa, and the ferromagnetism in $x$ = 0.16 appears to be completely suppressed. Our data clearly show that this system can be brought to a QPT and, hopefully a QCP. It will be very interesting to study the compounds via transport measurements under pressure and evaluate their critical exponents at $p_{\rm c}$. In addition, alternative methods of growing higher $x$ compounds or pressure studies on pure LaCrGe$_3$ will be possible ways to tune the potential QCP system as well.

\begin{acknowledgments}
We thank W. E. Straszheim for his assistance with the elemental analysis of the samples. This work was carried out at the Iowa State University and supported by the AFOSR-MURI grant No. FA9550-09-1-0603 (X. Lin, V. Taufour and P. C. Canfield). S. L. Bud'ko was supported by the U.S. Department of Energy, Office of Basic Energy Science, Division of Materials Sciences and Engineering. Part of this work was performed at Ames Laboratory, US DOE, under Contract No. DE-AC02-07CH11358. X. Lin and P. C. Canfield also acknowledge H. Miyazaki for key inspiration. 
\end{acknowledgments}


\begin{thebibliography}{00}

\bibitem{Schlenker_2005}
$\acute{E}$. du Tr$\acute{e}$molet de Lacheisserie, D. Gignoux and M. Schlenker (Eds.), \textit{Magnetism: Fundamentals}, (Springer, Boston, 2005).

\bibitem{Uhlarz_2004}
M. Uhlarz, C. Pfleiderer and S. M. Hayden, Phys. Rev. Lett. 93, 256404 (2004).

\bibitem{Thessieu_1995}
C. Thessieu, J. Flouquet, G. Lapertot, A. N. Stepanov and D. Jaccard, Solid State Comm. 95, 707 (1995).

\bibitem{Stoner}
E. C. Stoner, Phil. Mag. 15, 1018 (1933).

\bibitem{Sachdev_1999}
S. Sachdev, \textit{Quantum Phase Transitions} (Cambridge University Press, Cambridge, England, 1999).

\bibitem{Lonzarich_1986}
G. G. Lonzarich, J. Magn. Magn. Mater. 54, 612 (1986).

\bibitem{Lonzarich_1988}
G. G. Lonzarich, J. Magn. Magn. Mater. 76, 1 (1988).

\bibitem{Hertz_1976}
J. Hertz, Phys. Rev. B 14, 1165 (1976).

\bibitem{Stewart_1984}
G. R. Stewart, Rev. Mod. Phys. 56, 755 (1984).

\bibitem{Stewart_2001}
G. R. Stewart, Rev. Mod. Phys. 73, 797 (2001).

\bibitem{Stewart_2006}
G. R. Stewart, Rev. Mod. Phys. 78, 743 (2006).

\bibitem{Bud'ko_2004}
S. L. Bud'ko, E. Morosan and P. C. Canfield, Phys. Rev. B 69, 014415 (2004).

\bibitem{Bud'ko_2005}
S. L. Bud'ko, E. Morosan and P. C. Canfield, Phys. Rev. B 71, 054408 (2005).

\bibitem{Mun_2013}
E. D. Mun, S. L. Bud'ko, C. Martin, H. Kim, M. A. Tanatar, J.-H. Park, T. Murphy, G. M. Schmiedeshoff, N. Dilley, R. Prozorov and P. C. Canfield, Phys. Rev. B 87, 075120 (2013).

\bibitem{Taufour_2010}
V. Taufour, D. Aoki, G. Knebel and J. Flouquet, Phys. Rev. Lett. 105, 217201 (2010).

\bibitem{Saxena_2000}
S. S. Saxena, P. Agarwal, K. Ahilan, F. M. Grosche, R. K. W. Hasselwimmer, M. J. Steiner, E. Pugh, I. R. Walker, S. R. Julian, P. Monthoux, G. G. Lonzarich, A. Huxley, I. Sheikin, D. Braithwaite, and J. Flouquet, Nature (London) 406, 587 (2000).

\bibitem{Flouquet_2001}
A. Huxley, I. Sheikin, E. Ressouche, N. Kernavanois, D. Braithwaite, R. Calemczuk, and J. Flouquet, Phys. Rev. B 63, 144519 (2001).

\bibitem{UCoGe}
N. T. Huy, A. Gasparini, D. E. de Nijs, Y. Huang, J. C. P. Klaasse, T. Gortenmulder, A. de Visser, A. Hamann, T. G$\ddot{\rm o}$rlach and H. v. L$\ddot{\rm o}$hneysen, Phys. Rev. Lett. 99, 067006 (2007).

\bibitem{Appel_1980}
D. Fay, and J. Appel, Phys. Rev. B 22, 3173 (1980).

\bibitem{Varma_1986}
K. Miyake, S. Schmitt-Rink, and C. M. Varma, Phys. Rev. B 34, 6554 (1986).

\bibitem{Fisk_2006}
D. A. Sokolov, M. C. Aronson, W. Gannon and Z. Fisk, Phys. Rev. Lett. 96, 116404 (2006).

\bibitem{YbRh2Si2}
O. Trovarelli, C. Geibel, S. Mederle, C. Langhammer, F. M. Grosche, P. Gegenwart, M. Lang, G. Sparn and F. Steglich, Phys. Rev. Lett. 85, 626 (2000).

\bibitem{Mar_2007}
H. Bie, O. Y. Zelinska, A. V. Tkachuk and A. Mar, Chem. Mater. 19, 4613 (2007). 

\bibitem{Avdeev_2013}
J. M. Cadogan, P. Lemoine, B. R. Slater, A. Mar and M. Avdeev, Solid State Phenom. 194, 71 (2013).

\bibitem{Mar_2009}
H. Bie and A. Mar, J. Mater. Chem. 19, 6225 (2009). 

\bibitem{Binary}
T. B. Massalski, \textit{Binary Alloy Phase Diagrams}, 2nd ed., (ASM International, New York, 1990).

\bibitem{Canfield_1992}
P. C. Canfield and Z. Fisk, Philos. Mag. B 65(6), 1117 (1992).

\bibitem{Canfield_2010}
P. C. Canfield, Solution growth of intermetallic single crystals: a beginners guide, in: E. Belin-Ferr$\acute{\rm e}$ (Eds.), \textit{Properties and Applications of Complex Intermetallics}, (World Scientific, Singapore, 2010).

\bibitem{X-ray}
C. J. Howard and B. A. Hunter, \textit{A Computer Program for Rietveld Analysis of X-Ray and Neutron Powder Diffraction Patterns}, (NSW, Australia: Lucas Heights Research Laboratories, 1998).

\bibitem{Pressure}
P. L. Alireza, S. Barakat, A. M. Cumberlidge, G. Lonzarich, F. Nakamura and Y. Maeno, J. Phys. Soc. Jpn. 76SA, 216 (2007). 

\bibitem{Arrott}
A. Arrott, Phys. Rev. 108, 1394 (1957).

\bibitem{Noakes}
A. Arrott and J. E. Noakes, Phys. Rev. Lett. 19, 786 (1967).

\bibitem{Williams_1986}
I. Yeung, R. M. Roshko, and G. Williams, Phys. Rev. B 34, 3456 (1986).

\bibitem{Jia_2008}
S. Jia, N. Ni, G. D. Samolyuk, A. Safa-Sefat, K. Dennis, H. Ko, G. J. Miller, S. L. Bud'ko, P. C. Canfield, Phys. Rev. B 77, 104408 (2008).

\bibitem{Moriya_1973}
T. Moriya and A. Kawabata, J. Phys. Soc. Jpn. 35, 669 (1973).

\bibitem{Huber_1975}
J. G. Huber, M. B. Maple, D. Wohlleben and G. S. Knapp, Solid State Comm. 16, 211 (1975).

\bibitem{RWR}
P. Rhodes and E. P. Wohlfarth, Proc R Soc Lon Ser-A 273, 247 (1963).

\bibitem{Mydosh}
J. A. Mydosh, \textit{Spin Glasses: An Experimental Introduction}, (Taylor $\&$ Francis, London, 1993)


\end{thebibliography}
\end{document}